# *Wherefore Art Thou?* Provenance-Guided Automatic Online Debugging with Lumos

Jingyuan Chen § Lei Zhang † Leon Schuermann § Gongqi Huang § Ravi Netravali § Amit Levy §
§ *Princeton University* † *ByteDance*

## Abstract

Debugging distributed systems in-production is inevitable and hard. Myriad interactions between concurrent components in modern, complex and large-scale systems cause non-deterministic bugs that offline testing and verification fail to capture. When bugs surface at runtime, their root causes may be far removed from their symptoms. To identify a root cause, developers often need evidence scattered across multiple components and traces. Unfortunately, existing tools fail to quickly and automatically record useful provenance information at low overheads, leaving developers to manually perform the onerous evidence collection task.

Lumos is an online debugging framework that exposes application-level bug provenances—the computational history linking symptoms of an incident to their root causes. Lumos leverages dependency-guided instrumentation powered by static analysis to identify program state related to a bug's provenance, and exposes them via lightweight on-demand recording. Lumos provides developers with enough evidence to identify a bug's root cause, while incurring low runtime overhead, and given only a few occurrences of a bug.

## 1 Introduction

The growing complexity of modern cloud systems makes debugging more challenging than ever. While testing and verification [15] can catch large classes of bugs ahead of deployment, many bugs remain undetected until deployment. Bugs can result from intricate combinations of runtime conditions, including user inputs, ordering of concurrent events [24], and cross-component interactions [36]. Worse, a bug's root cause can be far from its manifestation, may span multiple components, multiple executions, and may only be reflected in state not captured in log outputs. To gather evidence that explains how a previous execution led to a bug, developers need to perform backward reasoning from the bug's manifestation, in the production system itself. We term this process *provenance-guided online debugging*.

Automated solutions should be able to resolve these issues, however current approaches fail to do so in one of three ways: (1) They incur impractically high runtime overhead by recording too much information [31, 32]. (2) They collect insufficient information to reliably identify root causes in many cases [22]. (3) They require too many iterations to collect the *right* information [1, 19, 34, 44].

In order for an automatic online debugging tool to be *practical*, it must satisfy three criteria concurrently:

| | RR | Sampling | Static Log | **Lumos** |
|---|---|---|---|---|
| Low Overhead | | ✓ | ✓ | ✓ |
| Diag. Latency | ✓ | | ✓ | ✓ |
| Soundness | ✓ | Eventually | | ✓ |

Table 1: Existing approaches to online debugging, including record-replay (RR), sampling, and static logging each fail to be practical in at least one of three ways: low runtime-overhead, low diagnostic latency, or soundness.

1. **Low Runtime Overhead**: Since online debugging instruments a system in-production, it should impose low performance overhead.
2. **Low Diagnosis Latency**: It should take few occurrences of the bug to collect enough evidence to start performing root cause reasoning.
3. **Soundness**: Developers should be able to trust that the results of a debugger contain enough evidence to correctly identify the root cause.

We propose Lumos, an online debugging tool for automated evidence collection in distributed systems that achieves all three. With Lumos, a developer specifies the *manifestation* of an issue to investigate—such as a panicking line of code—and Lumos *dynamically* instruments tracepoints that capture and correlate evidence for recovering the provenances of bugs.

Unlike prior approaches, Lumos achieves all three critical properties (Table 1) using two key techniques: *dependency-guided instrumentation* and *lightweight data flow tracing*.

*Dependency-guided instrumentation* allows Lumos to automatically choose provenance-relevant tracepoints to be instrumented, powered by a scalable static analysis that estimates potential data and control dependencies among complex system components. Under this strategy, Lumos is able to adjust its instrumentation based on the bugs under investigation, and avoids overheads introduced by collecting irrelevant information.

With the tracepoints identified, Lumos' *lightweight data flow tracing* makes use of a set of inexpensive tracing techniques for correlating data flows caused by accesses to shared state. This allows Lumos to associate evidence scattered in multiple traces to piece together complex provenances of bugs.

We evaluate Lumos on two representative real-world bugs found in the large, complex, and widely-deployed Hadoop Distributed File System. Our evaluation focuses on particu-larly intricate bugs that complicate root-cause reasoning. Our experiments demonstrate that Lumos incurs practical runtime overheads (approx. 10% reduction in throughput at high loads on average) during an active debugging session, and identifies root causes within 5 occurrences.

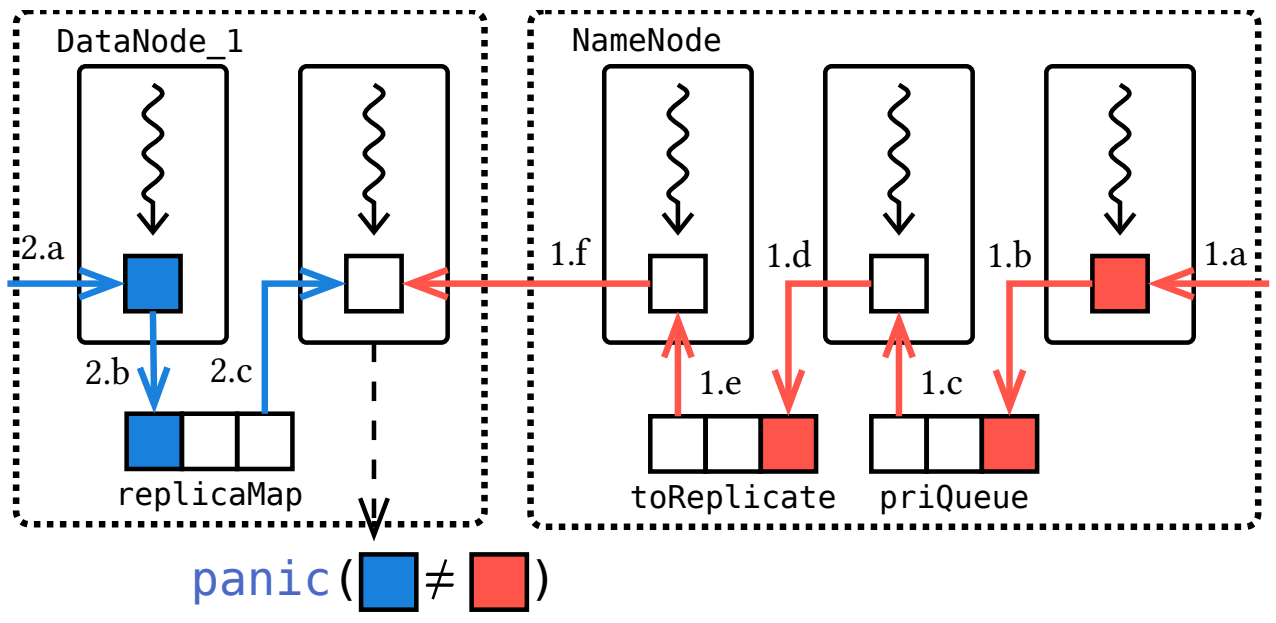

Figure 1: Development Process of HDFS-4022

## 2 Motivation

Bugs in large-scale distributed systems can be complicated. Recent works have observed that many distributed system bugs are non-deterministic and environment-dependent, such as configuration errors [17], failure-induced bugs [18], concurrency-related bugs [24, 26], or cross-component interactions bugs [36]. As it is infeasible to model the entire space of possible runtime states and environments ahead of time, simple pre-production testing will inevitably fail to uncover some of these bugs [15]. Therefore, developers have to troubleshoot them after they manifest during deployment.

Yet the multi-component and stateful nature of distributed systems make this process challenging. The pieces of evidence for reasoning about their root cause may be *latent*, meaning they can be far from a bug's manifestation (both in time, and in distance between the code fragments), and the evidence linking them may be scattered across multiple components and traces. To locate the root cause, debuggers need to perform backward reasoning to uncover the previous computational steps leading to the bug, beyond the information available at the point where a bug manifests (e.g., a `panic()`). We refer to this process as *provenance-guided debugging*. We next illustrate the necessity of such a process through a real-world bug HDFS-4022 [2].

### 2.1 Troubleshooting a Latent and Non-Deterministic Bug in HDFS

The Hadoop Distributed File System (HDFS) is a production-scale distributed filesystem, consisting of two types of components: one main `NameNode` responsible for managing file metadata, and multiple `DataNodes` responsible for persisting and replicating the actual file blocks. `Clients` may communicate with the `NameNode` to obtain or update metadata of files, with which they can communicate with `DataNodes` to transfer the file contents. The `DataNodes` communicate with the `NameNode` by periodically sending heartbeats and reports to indicate their liveness and work progress. Each `DataNode` may also transfer blocks to other `DataNodes` for replication.

Figure 1 illustrates how the HDFS-4022 [2] bug manifests in this system. For this bug to occur, a `Client` first writes to a file and update the last block's metadata by calling `NameNode:completeFile()` (1.a). This results in a `Block` (red) to be added to the `priQueue` data structure at the `NameNode` (1.b). Later, a background thread retrieves it (1.c) and add it to the `toReplicate` queue (1.d). Next, the client calls `append()` on the file, which results in a new version of the `Block` (blue) being generated. This `Block` is sent to `DataNode_1` (2.a) and stored in its `replicaMap` (2.b). Later, `DataNode_2` joins the system, which triggers the `NameNode` to ask `DataNode_1` to replicate the block (1.f). However, the `NameNode` retrieves the outdated `Block` when sending its metadata to the `DataNode_1` (1.e). `DataNode_1` then compares the received `Block` with the matching `Block` from `replicaMap` (2.c), and reports an error due to mismatches in their versions.

The root cause was a missing delete of the outdated `Block` (red) from `toReplicate` after `append()` takes place. Tracking it down requires developers to backtrack the provenance of the two blocks all the way to Step 1.a and 2.a to conclude that `Block` (red) is indeed outdated. The entire chain of evidence spans two components, four traces and three shared data structures, which poses great challenges for developers to capture it in its entirety. Even worse, the bug only happens under rare interleavings of events, such as the late joining of `DataNode_2`.

This example illustrates key challenges in debugging distributed system incidents, whose manifestation and root causes can be fairly simple, but the chains of propagation are complex. Such complexities are common in real-world incidents [3, 4].

### 2.2 Provenance-Guided Debugging Today

Next, we briefly review how developers collect provenances of in-production bugs today. One of the simplest and most common techniques is logging. However, placement of logging statements are usually pre-determined and only record a limited amount of information. This is insufficient and brings difficulty for root cause reasoning when additional bug-specific evidence is needed.

Another option is to record everything at runtime (similar to record-and-replay tools [31]). This option can recover arbitrary information developers may need, but at the cost of prohibitive runtime overheads.

Finally, recent advances in failure reproduction tools allow developers to debug in a synthesized offline environment produced by symbolic execution [45, 47]. However, such tools suffer from scalability and cost issues when handling complex bugs. Moreover, these tools only guarantee to reproduce the failure manifestation with no promises to faithfully recover the original root causes.

All of the aforementioned practices fail to strike a good balance between runtime overheads and universal bug-specific evidence collection. To resolve this tension, sampling-based recording may be used, where each execution trace is selected at random to be recorded according to predetermined sampling rates. Yet the sampling rates are usually kept very low (around 1%) to meet the goal of low overheads, resulting in collected evidence being fragmented and sparse. Debuggers may need to wait until they are lucky where a complete chain of evidence is sampled, leading to slow diagnosis.

### 2.3 The Need for Automatic, Efficient, and Flexible Online Debugging

In order to pinpoint root causes in a timely fashion, developers must reason backward from where the bug manifests and add logs or tracepoints to observe the values of relevant variables that explain the bug's provenance. Unfortunately, all of the previous techniques either incur high overheads that are intolerable for production systems, or capture insufficient information, making provenance-guided debugging a time-consuming and manual process. And worse, for any practical systems of significant size, performing this process *manually* is implausible, since the number of potentially relevant variables grow exponentially as the chain of evidence deepens. Therefore, we need a debugging tool that automates the process of bug-specific provenance recording, with as few manual steps and as little overhead as possible.

## 3 Lumos to the Rescue

To enable fast provenance-guided debugging at low runtime overheads, we present Lumos, a debugging tool that automatically instrument logs and tracepoints to collect provenance-related evidence for runtime bugs. When a bug occurs, developers simply specify how the bug manifests (e.g., through a particular value of `x` in a `panic(x)`) to Lumos. Lumos then performs *dependency-guided instrumentation*, under which values of variables that the bug manifestation has data/control dependency on are selected to be recorded. Such a *systematic* selection process avoids overheads incurred by recording provenance-irrelevant information and ensures sufficient necessary evidence is recorded. Compared with sampling, it can recover provenances with much fewer occurrences of the bug, leading to significantly lower diagnosis latencies.

Lumos' automatic instrumentation is powered by a scalable static analyzer that models all *application-space* control and data dependencies, including heap data dependencies and dependencies across multiple functions and threads. This empowers Lumos to select any provenance-relevant computation steps in complex code bases. Lumos assumes that library functions are correct and therefore only collects application-space information. We argue that this design is reasonable under the observation that most real-world application bugs can be troubleshooted through application-space evidence. However, we stress that Lumos is *capable* of modeling application-space dependencies *induced* by library functions. We discuss this distinction in detail in Section §3.2.

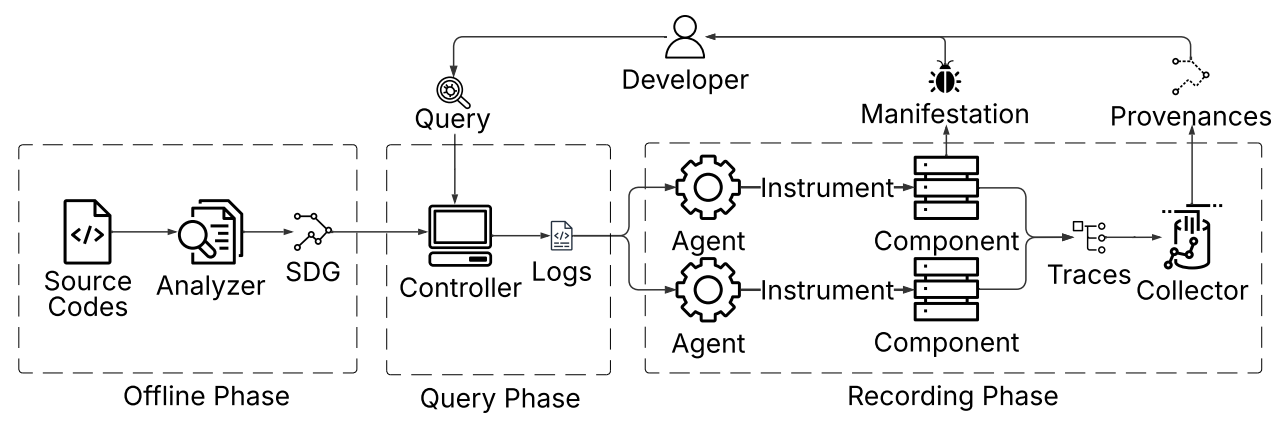

Figure 2: Lumos Architecture.

### 3.1 The Lumos System Architecture

Figure 2 illustrates Lumos' architecture and workflows. At the *offline phase* (Section §3.2), Lumos performs static analysis to build an application-space *system dependency graph* (SDG). In an SDG, each node represents an access to a memory location (local variable or heap location), and each edge stands for a static dependency between two accesses. The SDG is loaded into a centralized *controller*. When a bug occurs, Lumos enters the *query phase* (Section §3.3). The developers first collect their current knowledge about the bug. When the bug occurs for the first time, this includes the manifestation point (e.g., where a `panic()` happens). When a bug reoccurs, this information may be augmented with traces and partial provenances exposed by logs instrumented by Lumos from previous occurrences.

This knowledge is then submitted as *debugging queries* to the controller, which generates an *instrumentation plan* based on the queries and the SDG. The plan specifies recording operations that capture runtime values of nodes potentially relevant to the provenance of the bug (up to a tunable overhead limit). These operations are sent to the Lumos agents running alongside each system component and are then instrumented by the agents at runtime. Lumos then transitions to the *recording phase* (Section §3.4) where the agents continuously generate traces. When the bug reoccurs, the agents collect and send the traces to a centralized collector. Developers then assemble the provenances for root-cause-reasoning and potentially formulate queries for *futher rounds* to collect additional evidence if the root cause cannot be deduced.

**Challenges.** Achieving low diagnosis latencies and runtime overheads with the above architecture comes with the following challenges. First, Lumos' analyzer must strike a good trade-off between precision and scalability when identifying dependencies. This is crucial for Lumos to avoid recording excessive amounts of data. Past works on static

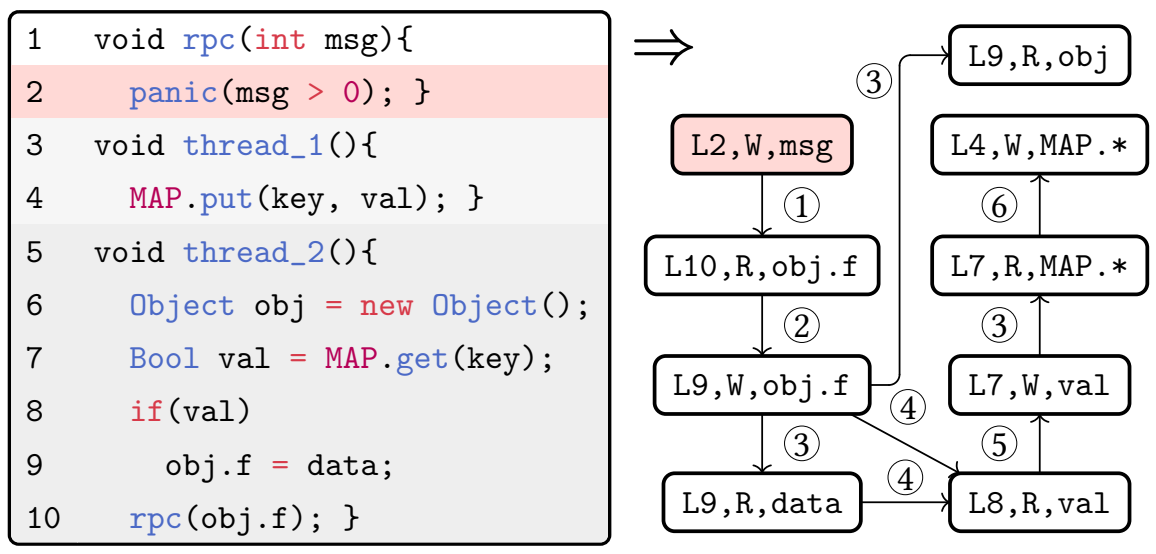

Figure 3: An example of SDG. Accesses to reference (s,f) is abbriviated as s.f.

analysis [22, 23] have identified this to be difficult on large code bases.

Second, given the selected nodes that are considered bug-provenance-relevant, Lumos' controller must generate a recording plan that traces the values of and data flows between the nodes with as few logs as possible while avoiding expensive recording operations.

And third, Lumos' agents must carry out the recording plan efficiently with lightweight recording operations, while maintaining negligible overheads when no debugging is in-progress. We specify how Lumos addresses the challenges throughout the remainder of this section.

## 3.2 Offline Phase: Static Analysis

To enable fast diagnosis with low runtime overheads, Lumos must respond swiftly to developers' queries and automatically identify as small a set of bug-provenance-relevant recording operations as possible. To this end, Lumos performs offline static analysis that outputs data and control dependencies between accesses to variables and heap locations, such that dependent accesses can be extracted efficiently at runtime.

### 3.2.1 Definitions

**Statement and AML.** Lumos models program execution as atomic transitions between `Stmt`s. Each `Stmt` may access (read or write) a set of abstract memory locations (`AML`s). Lumos models two types of `AML`s: local variables (`Local`, e.g., `val` at Line 7 in Figure Figure 3) and heap references (`Reference`, e.g., `obj.f` at Line 9).

**System Dependency Graph.** Lumos' analyzer outputs dependency relations between any variables and heap locations accessed, captured in a so-called system dependency graph (SDG). In this graph, each `Node` corresponds to an access, defined as a tuple (`Stmt`, `Type`, `AML`) that specify the statement and the type (`R`/`W`) of the access, as well as the accessed `AML`. For example, at `L9` in Figure Figure 3, three nodes are constructed: `(L9,R,obj)`,`(L9,W,obj.f)` and `(L9,R,data)`. Each edge represents a control/data dependency relation between a pair of access nodes. We next specifies the different types of dependencies Lumos models and elaborate on how they are derived. An example code snippet and part of its SDG are illustrated in Figure Figure 3. Throughout the next sections we use $n$ to represent an SDG `Node`, and $s_n$ to represent its `Stmt`, $l_n$ for its `AML`, and $t_n$ for its `Type`.

```
1  void caller1(int y){
2    o1.f=y;
3    print(foo(o1));
4    bar(o1); }
5  foo(o){return o.f;}
6  bar(o){return o;}
```

```
7  void caller2(int z){
8    o2.f=z;
9    print(foo(o2));
10   o3=bar(o2);
11   print(o3.f); }
```

Figure 4: Examples of drops in analysis precision caused by the calling-context problem.

### 3.2.2 Control Dependency

A control dependency $n_1 \rightarrow n_2$ (such as ④ in Figure 3) exists if $l_{n_2}$ is a control variable at a branching statement $s_{n_2}$, and the reachability of $s_{n_1}$ depends the branch taken. Lumos performs post-dominator analysis [10] to identify control dependencies between statements. We curently do not support exceptional control dependencies. In addition, Lumos also resolves inter-procedural control dependencies via *call graph analysis* [9].

### 3.2.3 Data Dependency

A data dependency $n_1 \rightarrow n_2$ exists if changing the value of $l_{n_2}$ may affect that of $l_{n_1}$ according to the data flow semantics of the programming language under analysis. There are multiple types of data dependencies: an intra-statement data dependency ③ exists when $s_{n_1} = s_{n_1}, t_{n_1} = W, t_{n_2} = R$. The relation captures the fact that the computational effects of each statement depend on the inputs. Meanwhile, an inter-statement data dependency (②③⑤⑥) exists when $s_{n_1} \neq s_{n_2}, t_{n_1} = W, t_{n_2} = R$ and $l_{n_1}$ and $l_{n_2}$ may refer to the same memory location, which captures the fact that a read from a `AML` depends on a write to it. This can be further catagorized as follows:

**Local Data Dep.** When $l_{n_1} = l_{n_2} = V$, where $V$ is a `Local`, a local data dependency like ⑤ arises.

**Inter-Proc Dep.** An interprocedural data dependency ① arises due to passing data via parameters or return values. This happens when $n_1$ represents a read from a formal parameter of a method $m$, and $n_2$ represents the write to the actual argument passed at a callsite of $m$. The case for return values is similar. To obtain these facts, Lumos relies on the result of the call graph analysis that outputs pairs of callsites and target methods. An inter-procedural dependency may span multiple components, such as the dependency ① caused by an RPC in the example of Figure Figure 3. To add such dependencies edges, we first annotate the communication endpoints (call site or entrypoint of RPCs), then add dependency edges between the matching nodes at each side of the

```
1  List l1.add(o1);
2  List l2 = l1;
3  o2 = l2.get(i);
```

```
4  URI u1 = new URI(..);
5  URI u2 = u1.resolve(..);
6  File f = new File(u2);
```

Figure 5: Examples illustrating the analysis of library-induced dependencies.

communication channel, including pairs of formal parameters and arguments, return values and corresponding data structures in the serialization and deserialization functions.

**Heap Data Dep.** A heap data dependency arises when $l_{n_1} = A_1, l_{n_2} = A_2$, and $A_1$ and $A_2$ are `References` that may refer to the same heap location. Lumos handle application ② and library ⑥ heap data dependencies separately. Each `Reference` is modeled as a pair of (`AML`, `Field`), where `AML` describes the reference's base (which can be another `Reference` or a `Local`), and `Field` describes the field accessed relative to its base. For example, for an access to `s.f`, the `Reference` is `(Local s, f)`.

A pair of application `References` $A_1 = (b_1, f_1)$ and $A_2 = (b_2, f_2)$ may refer to the same location if $f_1 = f_2$ and $b_1$ *may alias* with $b_2$. Lumos obtains this information through Anderson-style points-to analysis [29]. When a heap dependency involves a `Reference` to a data structure shared between threads (e.g., ⑥), Lumos additionally refers to it as a shared state dependency. Shared data structures are identified via thread escape analysis [33].

#### 3.2.4 Challenges in Precision vs. Scalabililty

Lumos requires precise heap dependency analysis to avoid large amounts of spurious dependency edges in its SDGs, which would lead to excessive amounts of data logged in the query phase. Unfortunately, traditional heap dependency analysis can be imprecise [22, 23] for complex code bases. This is due to the effect of calling contexts as illustrated in Figure 4: `foo()` has two *calling contexts* (call sites) `L3` and `L9`. To distinguish between dataflow facts involving the effect of `foo` under different contexts, such as whether `(L3,foo(o1))` depends on `(L2,y)`, a *context-sensitive* analysis may be used with which different copies of `foo` under each context are created and analyzed separately. Yet such context-sensitive analysis is hard to scale due to the exponential growth in the number of contexts as the call chains deepen [25, 35]. Hence, practical analysis must sacrifice precision by adopting *context-insensitivity* that does not distinguish between contexts. This results in two types of imprecision.

**Type 1.** False-positive heap dependency edges may arise due to noisy aliasing pairs. In the above example, a noisy aliasing pair `(o1,o3)` will be generated due to the two contexts of `bar()` at `L4` and `L10`, resulting in a spurious edge `(L11,R,(o3,f))`→`(L2,W,(o1,f))`.

**Type 2.** False-positive *paths* may occur when callees with multiple callsites are involved, like the path `(L3,R,foo(o1))` →`(L5,R,(o,f))`→`(L8,W,(o2,f))`.

Lumos resolves this fundamental tension between the analysis' scalability and precision with the following design: for Type 1 imprecision, Lumos uses a state-of-the-art points-to analysis [29] that avoids a large portion of context-induced precision drops when analyzing aliasing relationships. The algorithm "short-cuts" the analysis of points-to relations for heap-manipulating functions whenever possible without incurring the complexity of context-sensitive algorithms.

For Type 2 imprecision, Lumos avoids constructing edges from/to nodes within functions with many calling contexts as much as possible. Two sub-designs enable this. First, Lumos inlines short functions that solely perform heap accesses. This helps to avoid noisy paths caused by short but frequently-called functions (e.g., getters and setters like `foo` above). As the functions are short, copying them at each callsite only increases the graph size by a small amount. Second, Lumos utilizes an observation from recent work [11, 20] that the imprecision in heap dataflow analysis is mostly caused by library functions with complex internal dynamics. However, the externally observable semantics of these library functions can be much simpler. Since Lumos only tracks application-space provenances, it is possible to bypass the internals of complex library functions and derive the effects of library calls to them based on their externally-visible semantics. Below we elaborate on this.

**Library-Bypassed Analysis.** Lumos employing ideas from recent works in static analysis [11, 12] on bypassing the internals of library functions. At each library callsite, Lumos generates *specifications* that summarize the *side effects* of the calls. For each `s:Local` of non-premitive types used at a callsite (arguments or return values), the specifications tell whether `(s,*)` is read or written. where * is a *pseudo field* that represents the set of all *reachable heap locations* from the base reference. Lumos then generates the access nodes based on the specifications. For example, for callsite `L4` in Figure Figure 3, `(L4,W,(MAP,*))` is generated.

To determine whether `(L1,R,(s1,*))` depends on `(L2,W,(s2,*))`, one must check whether `(s1,*)` and `(s2,*)` *may overlap*. If `s1` may-alias with `s2`, then they *directly* may-overlap. For instance, `(L3,R,(l2,*))` depends on `(L1,W,(l1,*))` in Figure 5 as `l1` may-alias with `l2`. Otherwise, Lumos performs additional analysis to determine potential *indirect* may-overlap relations. Lumos conservatively deems `(s1,*)` and `(s2,*)` to overlap indirectly if 1) `s1` and `s2` appears at a common callsite, and at least one of `(s1,*)` and `(s2,*)` is written; 2) there is an `s3` that both `s1` and `s2` overlap with. For example, in Figure Figure 5, `(u1,*)`, `(u2,*)`and `(f,*)` pair-wise may-overlap. To prevent significant precision drops, Lumos only tracks indirect overlapping between library objects within the same function and ignores inter-procedural cases. While this may cause missed dependencies in edge cases, we argue that these cases are rare; in all our case studies, we did not encounter crucial library dependencies to be absent.

Lumos uses templates to efficiently generate the speficiations from a small set of developer provided annotations that specify the rules for the generation, which only involves 517 LoC in Datalog, plus a short list of 112 functions to be treated by special rules. Automatically generating such semantics an active area of research [12, 20, 38]; we leave this extension to future work.

### 3.3 Query Phase: Inst-Plan Generation

When a bug occurs, a developer collects their current knowledge about the bug and presents it as a set of *debugging queries* to the Lumos controller. For example, as the bug of Figure Figure 3 manifests at `L2: panic(msg>0)`, the query would be the node `(L2,R,msg)`. Given these queries, the Lumos controller is responsible for generating an instrumentation plan that specifies the recording operations to instrument. The plan must achieve two goals: first, it should ensure coverage of key values and data flows to enable reasoning about bug provenances. Second, it should avoid recording excessive amounts of information and using expensive recording operations.

To meet these goals, the controller selects nodes that the queried nodes depend on by performing a depth-limited traversal over the SDG, starting from the nodes in the queries. This algorithm works similar to program slicing [39]. The depth limit is set to avoid high recording overheads, as unbounded depth may result in all nodes selected. A larger depth could result in reduced diagnosis latency (the number of debugging rounds) as more nodes' values are recorded, at the cost of higher runtime overheads. Next, the controller identifies the necessary operations to be instrumented, namely ones responsible for recovering the values of nodes in $R$.

A brutal-force strategy could decide to record all values of nodes in $R$ and an ordering of all runtime events, which results in prohibitive overheads. Therefore, Lumos adopts *non-determinism-based recording* commonly used in record-and-replay frameworks [16, 40], under which only the sources of non-determinism are recorded such that the remaining values can be deduced from them. This includes the common environmental and input non-determinism: when $l_n$ is the return value of a non-deterministic library call, such as `System.nanoTime()`, the value is logged as the effect of the calls are not reproducible; when a node reads from an external input at the system boundary (such as an RPC entrypoint), the value is input-nondeterministic and logged. We currently annotate sources of environmental non-determinism, which involves 96 lines of codes in Datalog. In addition, we apply an optimization where only environmental/input non-determinism affecting nodes in $R$ is logged.

However, adopting the above strategy results two sub-challenges in the precense of thread-shared state. First, traditional RR frameworks record a *fixed* set of non-determinism at system boundaries, such as inputs and system calls, since they are *always-on.* This is insufficient for Lumos since recording in Lumos is enabled *on-demand.* As a result, the state of shared data structures when recording is enabled is an additional source of non-determinism from Lumos' perspective, as their values may be determined before recording was enabled. A naive solution is to take a heap snapshot when enabling recording, which may introduce long pauses in system execution that are impractical for in-production systems. To resolve this issue, Lumos *lazily* logs values read for each node in $R$ reading a shared `AML`.

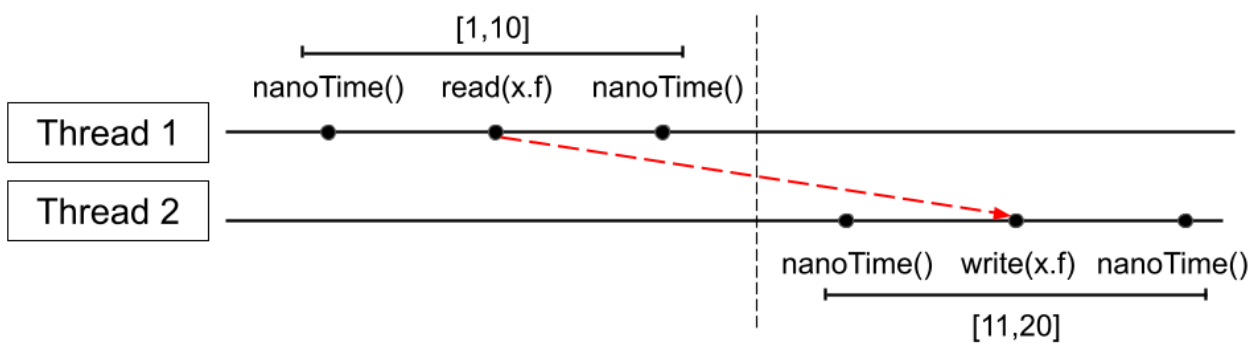

Figure 6: Example of using timestamps to approximate shared state accesses' orderings

Second, to recover bug provenances involving shared state, Lumos must resolve inter-thread dataflows (i.e., identifying the corresponding write event for each read event). This requires ordering information of shared state accesses. Yet traditional mechanism for recording ordering information use expensive mechanisms such as Lamport clocks[14] that result in high overheads. To tackle this problem, Lumos adopts the following technique.

#### 3.3.1 Lightweight Shared-State Dataflow Tracing

To address the second challenge, Lumos applies the *cross-interleave hypothesis* (CIH) from [21], which argued that when accesses to shared state are coarsely-interleaved, one may use the timestamps from shared hardware clocks to establish an approximation of their orderings. By assuming the validity of CIH, Lumos uses a set of *lightweight dataflow tracing operations* for shared state accesses. First, for each shared-state accessing node, Lumos logs the variables at the node whose values help to disambiguate the effect of the access at runtime. This includes the address of the base of the accessed `AML`, plus additional *witnesses* for arrays and collections. For instance, the index of the array element is logged as the witness for array accesses. For collections, the value of the added/retrieved element is logged as witnesses.

Second, the starting and ending timestamps of an interval containing the operations are logged. These intervals serve as an approximation of the orderings of the accesses. Figure 6 shows an example. A read and a write to `x.f` are from two threads. A pair of timestamps in each thread marks a temporal interval during which the access happened. When the intervals are disjoint and no other intervals overlap with them, the ordering of the accesses can be derived. Therefore, the inter-thread dataflow can be identified, as illustrated by the red dashed arrow.

This design does not guarantee to recover the exact ordering information, which may cause ambiguities when identifying inter-thread dataflows. When the intervals overlap, developers need to consider more than one ordering possibilities. However, we argue that CIH holds for most real-world distributed systems that do not perform frequent and highly concurrent accesses to *the same* shared variables. Therefore, the frequency of overlapping intervals should be low. In addition, we argue that developers can tolerate a small amount of ambiguities when debugging, and can further reduce the ambiguities via thread schedule synthesis tools like [43] to find a valid bug-triggering order of accesses. Furthermore, while the lack of exact ordering may introduce ambiguities, it is still sound and *does not* miss potential read-write pairs.

### 3.4 Recording Phase: Agent Operations

The Lumos agents are responsible for realizing the instrumentation plan while meeting two goals: first, they must ensure negligible overheads when recording phase is *disabled*. Second, they must achieve low-overhead recording when the phase is active.

To meet the first goal, the agents employ *dynamic instrumentation*. When recording is enabled, the agents insert logs at the corresponding statements, then compile and hot-reload the instrumented code. These operations are removed when exiting the recording phase, which incurs virtually zero overheads.

To achieve the second goal, Lumos utilizes a lightweight asynchronous tracing runtime. Each `log()` copies the payload to *thread-local* shared-memory buffers and returns immediately. The buffers' contents are asynchronously flushed to the centralized collector.

At the beginning of an entry point function, the agents insert a call to `startTrace()`, which obtains the thread-local buffers from a background buffer manager. Similarly, a call to `endTrace()` is inserted at the end that returns the buffers. Both calls also record timestamps marking the duration of the trace.

**Distributed Tracing.** In the presence of an RPC call, Lumos must link the caller's thread-local trace and the callee's trace into one coherent end-to-end trace, as dataflow may span both components. Lumos employs distributed tracing techniques to achieve this. Specifically, Lumos inserts a `setCaller()` right before each RPC call, which generates and logs a unique identifier for the call. The identifier is then embedded in the RPC's metadata and propagated to the callee's side, where the callee retrives and logs it via an inserted `getCaller()` at the beginning of the RPC entry function.

## 4 Implementation

We implement the Lumos analyzer with Doop, a Datalog-based static analysis framework, consisting of 2753 LoC in Datalog. The SDG construction and traversal algorithms are implemented in Neo4j with 343 lines of Cypher scripts and 115 lines of Datalog. For distributed tracing, Lumos uses the XTrace [13] framework. We developed the agent's instrumentation framework on top of Java Soot [37] with 7198 lines of Java. The framework uses the JVM's runtime bytecode reloading facility to achieve dynamic instrumentation, and employs Java's `System.nanoTime()` to record timestamps. For the asynchronous tracing runtime, we use the HindSight framework's tracing library [44] by wrapping it inside a JNI interface with 132 lines of Java and 126 lines of C.

## 5 Case Studies

In this section, we present case studies that illustrate how well Lumos enables provenance-guided debugging of complex issues. We use Lumos to debug two bugs from HDFS, a widely used distributed system. We chose the bugs based on the following criteria: first, these bugs demonstrates the complexity of *latent* root causes, where the evidence for root-cause reasoning are from multiple traces and components.

Second, the bugs are *non-deterministic*, which only occur under specific runtime conditions (e.g., concurrency, inputs, failures, etc.).

Third, we choose bugs with *reported* root causes, manifestations and backward reasoning processes. This provides us an *objective* standard to evaluate whether Lumos can record key evidence in online debugging scenearios.

**Methodology.** For each case study, we first reproduce the fault through a triggering workload. We then *simulate* a multi-round online debugging process that is *close* to the one from the original reports [41]. In each round, we formulate the debugging queries according to the report and the evidence collected so far to obtain the instrumentation plan. We tune the depth limit to keep the per-round overhead moderate. Our goal is to show that even with a depth limit that incurs only low and tolerable overheads in the range of 10%, Lumos can resolve complex, latent bugs in only a few, iterative, and highly automated debugging rounds. In reality, developers may choose a higher depth limit to reduce the number of rounds, at the cost of higher overhead. In addition, we keep the set of root-cause-relevant logs from previous rounds in the plan. We then inspect the traces to verify whether the key values and their provenances can be constructed, and visualize them as provenance graphs. We also qualitatively discuss why the *pre-determined logs* in the original code base failed to capture the key provenances.

As discussed in Section §3.3, Lumos only records sources of non-determinism. Therefore, to obtain the value of a specific variable, one may need to infer it from the other logged values. We currently perform this reconstruction process manually, and leave building an automatic replayer for the process as future work.

```
 1  void offerService(){ // DN
 2    List<Block> resp = sendHeartbeat();
 3    for(Block b : resp){
 4        ReplicaInfo ri = volumeMap.get(b.id);
 5        error(ri == null || b.GS != ri.GS); }}
 6  List<Block> sendHeartbeat(){ // NN
 7    while(...){pendingList.add(replicateBlocks.poll());}
 8    return pendingList; }
 9  void writeBlock(ReplicaInfo newRep){ // DN
10    volumeMap.put(newRep.id,newRep); }
11  int computeReplicationWork(){ // NN
12    while(...){
13      Block b = priQs.get(i).next();
14      replicateBlocks.offer(b); }}
15  void completeFile(INodeFile f){ // NN
16    for(Block b : f.getBlocks(){priQs.get(i).add(b); }}
```

Figure 7: HDFS-4022 relevant code fragments

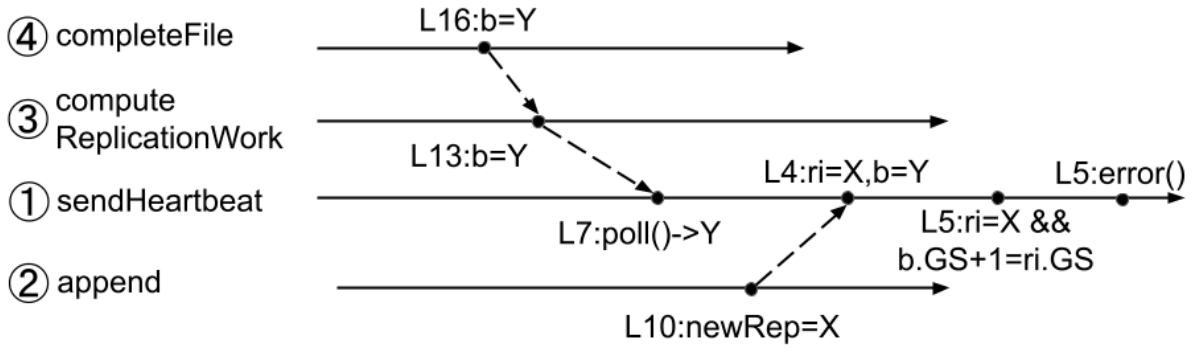

Figure 8: HDFS-4022 provenance graph

**HDFS-4022.** This is the motivational example from Section §2.1. Figure Figure 7 shows the simplified root-cause relevant code fragments, with the provenance graph shown in Figure 8. We annotate the *trace* that each statement or log belongs to in both figures. Each trace represents an end-to-end execution initiated by an external event. For example, the -`append` trace is initiated by user's request to append to a file. In addition, we annotate the component where each function executed. For this bug, the error manifests at `L5` reporting that for the incoming block to be replicated, either no replica information is found, or its timestamp does not match. Thus, the initial query is set to `(L5,R,ri == null || b.GS != ri.GS)`.

**Key Evidence 1.** Developers first find logs in trace showing `ri = X != null` and `b.GS + 1 == ri.GS` at `L5`. This indicates that the error was due to the `NameNode` asking the `DataNode` to transfer an outdated block. This evidence demonstrates Lumos' ability to record *concrete* values for developers to decide from which values to perform further backwards reasoning. Pre-determined logs lack this information, forcing developers to guess if the error was due to other conditions like `ri==null`.

**Key Evidence 2.** Developers then seek where `ri` comes from, and find it propagated from the shared `volumeMap` based on the logs `(L4:volumeMap=X,ri=Y)`. Seeking the corresponding write event that added `ri` to `volumeMap`, developers find a *potential* write from logs `(L10:volumeMap=X,newRep=Y)` in trace . Since no other potential writes are to be found, this event is considered as the *actual* writer. This evidence demonstrates Lumos' ability to track cross-trace shared-state provenances. With pre-determined logging, developers cannot find the write event that propagated `ri`.

**Key Evidence 3.** Based on the above, developers conclude that the problem lies in the block `b` instead of the replica information `ri`. They then seek where the block `b` at `L4` stems from. Based on the logs recording control-flows of the loop at `L3`, they understand that it orignates from `resp[i]`, where `i` is the loop iteration where the error at `L5` was reported. By backtracking further, they find that `resp[i]` was written by `pendingList` at `L7` in a different component (NN) under loop iteration `i`. This evidence shows Lumos' ability to track intra-trace provenances that potentially span multiple components. With predetermined logs, developers cannot figure out under which iterations the propagation `L7:pendingList[i]→L4:b` happened that led to the error at `L5` due to the lack of control-flow information and distributed tracing functionality.

**Key Evidence 4 & 5.** Developers then seek the origin of the block added at `L7`. At this point, no evidence in the current round is available. Therefore, developers start a second round with the query `(L7,R,poll())`, where `poll()` represents the call's return value. In the second round, they find that the block was propagated from `(L14,b)` in trace based on shared state logs similar to those in evidence 2. The provenance of this block is further found at `(L16,b)` in trace in a third round with query `(L14,R,b)`. While the last two pieces of evidence are simple, they are still difficult to pinpoint without the logs and traces due to the existence of other potentially relevant information. For example, `(L14,b)` may be propagated from a different trace than , under which the root cause may lie elsewhere.

**Root Causes.** At this point, developers can deduce the root cause. According to the semantics of the file system, `completeFile` was called the first time the client finishes modifying the block which adds the block to `priQs`. Later, the client appends to the file and indicates the `DataNode` that block has been modified by updating the `volumeMap`. However, the old block was not removed from the priority queue, causing the error.

**HDFS-5465&5479.** In this bug [5], certain blocks failed to be replicated. The developer starts debugging by noticing that the `NameNode` asked the `DataNode` to replicate an unexpectedly lower number of `Block` according to the reports. This bug spans two components and three threads. Additionally, this bug has two root causes, where one is a semantic misinterpretation and the other being a race condition. Figure 9 shows the relevant codes and Figure 10 shows the provenance graph. The initial query is set to node `(L6,R,(results,*))`.

**Key Evidence 1&2.** Developers first found that `results` is underpopulated due to the branch at `L5` not taken based

```
1   List<Block> sendHeartbeat(int xmitsInProgress){ // NN
2     int numTargets = maxRStreams-xmitsInProgress;
3     for(;!blockq.isEmpty() && numTargets>0){
4       numTargets-=blockq.peek().targets.length;
5       if(numTargets >= 0) results.add(blockq.poll()); }
6     return results; }
7   void offerService(){ // DN
8     for(Block b:sendHeartbeat(DN.xmitsInProgress)){..} }
9   void run(ReplicaInfo newReplica){ // DN
10    DN.xmitsInProgress++;
11    finally{DN.xmitsInProgress--;} }
```

Figure 9: HDFS-5465&5479 relevant codes

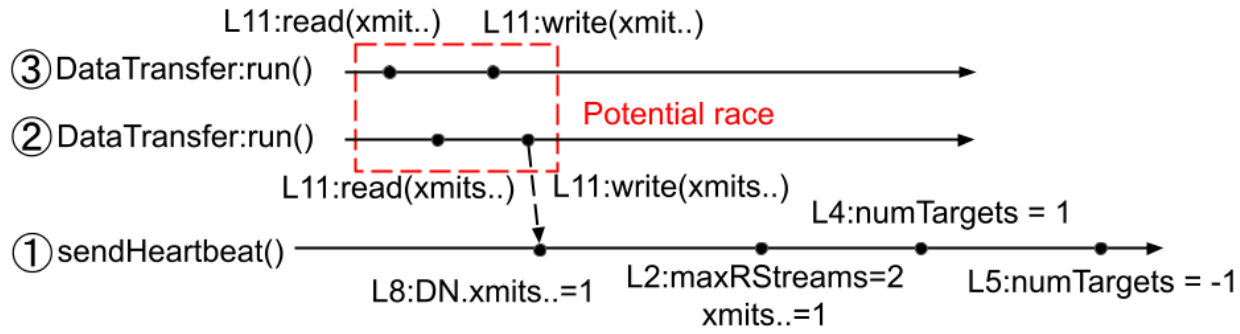

Figure 10: HDFS-5465&5479 provenance graph

on log (`L5,numTargets=-1`) in trace . This value was computed at `L4` where `numTargets=1` and `targets.length=2`. At this point, developers uncover the first root cause, which is the misinterpretation of `numTargets` as the number of blocks in the queue. Suspecting the initial value `numTargets=1` being problematic, developers seek and found its provenance at `L2`. The logs illustrates that `xmitsInProgress=1,maxRStreams=2`, causing `numTargets` to be lower than expected. This enables them to suspect a second root cause involving `xmitsInProgress`, which is the number of blocks currently being transmitting by the heartbeating `DataNode`. With pre-determined logging, developers cannot uncover the value of `numTargets` at the branch at `L6` as well as its provenances at `L5&L2`.

**Key Evidence 3.** Developers then start the next round of debugging with query (`L3,R,xmitsInProgress`). The traces illustrate that the value was propagated from the shared counter read (`L8,DN.xmitsInProgress=1`) in a different component (DN). With the available shared state logs, developers found multiple *potentially concurrent* writes to the counter at `L10&L11` in several traces such as and with overlapping intervals marked by the timestamps. With pre-determined logs, it can be challenging for developers to uncover the potential race.

**Root Causes.** Developers can now deduce that the faulty value of `xmitsInProgress` is *potentially* caused by a data race at `L10&L11`. While Lumos cannot unambiguously conclude the existence of data races, it still helps developers to trace the bug manifestation back to the anomolous shared variable.

# 6 Evaluation

In this section, we quantitatively evaluate how well Lumos enables fast diagnosis at low overheads.

## 6.1 Macrobenchmarks

We first conduct macrobenchmarks to evaluate whether Lumos achieves a good trade-off between diagnosis latency and overhead. We repeat the multi-step debugging process in Section §5 and measure the overhead added by Lumos' operations in each round, and count the number of rounds as the diagnosis latency. We then compare the results with that of the baselines.

**Setup.** We performed macrobenchmarks on a machine with 72 Intel Xeon Gold 5220 CPUs at 2.20GHz with 125G DRAM running Ubuntu 18.04. We used version 2.7.2 of Hadoop instrumented with XTrace [13] and modified the codes to inject the bugs. We run one `NameNode` and two `DataNodes`. We picked the benchmarks used in previous work [30], which includes 7 benchmarks from the NNThroughputbenchmark suites [6] for the `NameNode`, and a benchmark (read8k) for `DataNode` adapted from the DFSIO bench [7] that reads an 8K file. The benchmarks send requests from increasing numbers of close-loop clients in separate threads. For each experiment, 10000 samples are collected per thread.

**Baselines.** We compare Lumos against two baselines. The first is a *simulated* always-on record-and-replay (RR), where all operations are instrumented. This includes non-determinism logs, lightweight dataflow tracing operations and the remaining agent operations. To achieve a fair comparison, we do not instrument library functions. In addition, we excluded two subsystems in HDFS: the EditLog system (for persistence and recovery), and the data plane of the `DataNodes` (for reading and writing actual block contents). These components perform expensive I/Os and could result in high overheads when recording is enabled for them. Since none of our case studies require evidence in these components, having them instrumented could greatly increase the overhead of RR and would bias the benefits of Lumos.

The second baseline is sampled recording, which instruments the same amount of operations as RR, but only turns on recording for each trace with a sampling rate $p$. We estimate its diagnosis latency as $\left(\frac{1}{p}\right)^N$, which is the expected number of bug ocurrences needed before a full chain of provenance can be recorded. Here $N$ is the number of traces involved to pinpoint the root causes (e.g., $N = 4$ for HDFS-4022 and 3 for HDFS-5465). This exponentiation reflects that sampling is performed for each trace (or alternatively, at more fine-grained levels such as per-function) as a whole, and it requries all $N$ traces to be sampled to recover full provenances.

**Results.** We compare overheads of Lumos ($O_L$) to that of RR ($O_R$) with results summarized in Figure 15. The overheads are reported as the reduction in max throughput and the increase in average latencies at a low load (five concurrent clients), compared with the baseline with no Lumos' operations instrumented (only with the default logs and distributed tracing operations from the original system instrumented). The per-round overheads of Lumos for each case study are shown. For instance, L4022-1 represents the first round of case study HDFS-4022. Notice that Lumos' overheads are only incurred when it's in the recording phase, and are zero otherwise.

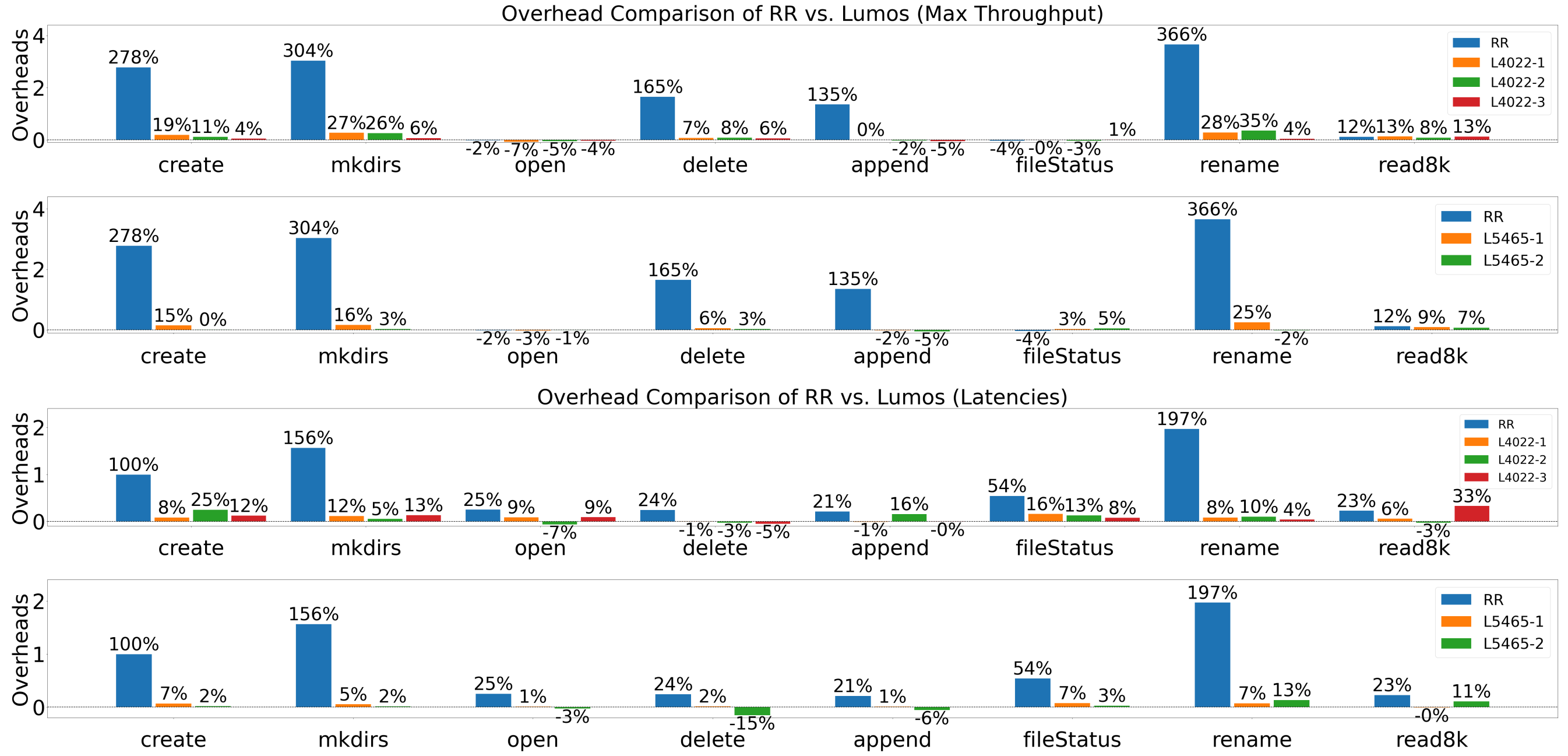

Figure 14: Overhead comparison for RR and Lumos. The relative overheads in terms of max throughput and average latencies at low load are shown for each case study separately. The x-axis labels represent the name of each benchmark. The first 7 benchmarks are the NameNode benchmarks, and last one (read8k) is the DataNode benchmark.

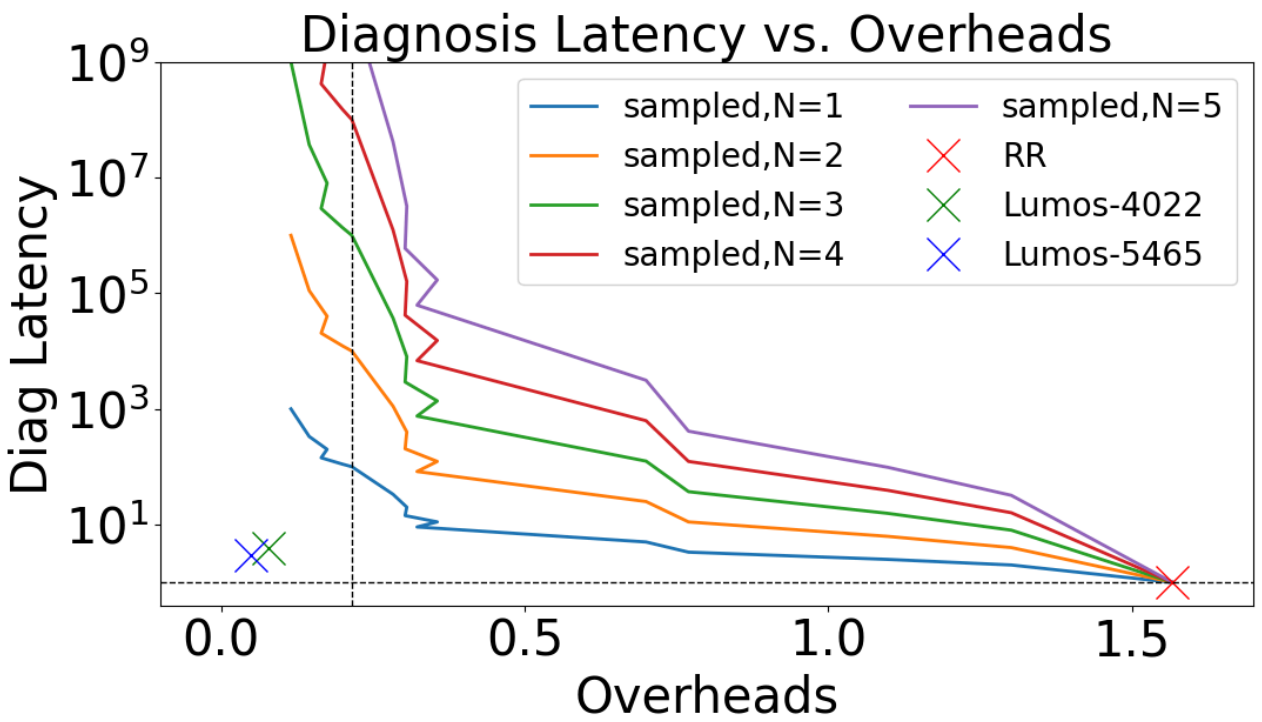

Figure 15: Diagnosis Latency vs Overhead comparison for RR, sampled recording and Lumos. Diagnosis latencies are shown in log scale. The vertical dash line shows the overheads for sampled recording with $p = 1\%$.

**Observation 1.** Lumos achieves practical per-round overheads (<11% reduction in max throughput on average across the benchmarks, and <9% in average latencies) that are *significantly* lower than RR.

Next, we compared the overhead-diagnosis-latency trade-offs of sampled recording, RR and Lumos with results shown in Figure 15. For sampled recording, we measure its overheads under different $p$s and plot the resulted diagnosis latency $\left(\frac{1}{p}\right)^N$ as a function of the overheads for different values of $N$. In addition, we mark the points representing RR and Lumos (for each case study). The diagnosis latency of Lumos is 4 reoccurrences for HDFS-4022 and 3 for HDFS-5465 as reported in Section §5. RR has a diagnosis latency of 1 since it records all information needed on the first bug occurrence. For Lumos and both baselines, we report the *average* overheads in max throughputs across the 8 benchmarks ($\overline{O_R} = 157\%$,$\overline{O_L} = 8\%$ across the rounds for HDFS-4022 and 5% for HDFS-5465). Notice that sampled recording incurs overheads all the time as it is *always-on*. Therefore, a practical sampling rate must be very low (such as 1%) to ensure the amortized overheads remain low.

**Observation 2.** Lumos' diagnosis latency is close to RR and several orders of magnitude lower than sampled recording under the same overhead budget. Combined with Observation, we conclude that Lumos not only achieves a good trade-off between overheads and diagnosis latencies, but also advanced the *pareto frontier* of the trade-off between the two goals.

| Agt. Op. | `startTrace` | `endT.` | `setCaller` | `getC.` |
|---|---|---|---|---|
| Lat. (ns) | 2606 | 678 | 1580 | 798 |

Table 3: Latencies of Agent Operations. Address is a long integer for the address of a Java object retrived from `System.identityHashCode()`. Str. is a Java string of length 10. Array is a byte array of length 32.

| Case | Analysis | Controller | | | Inst. (NN) | | | Inst. (DN) | | |
|---|---|---|---|---|---|---|---|---|---|---|
| | | Rd. 1 | 2 | 3 | 1 | 2 | 3 | 1 | 2 | 3 |
| 4022 | 25m34s | 122 | 15 | 10 | 9 | 11 | 11 | 13 | 1 | 1 |
| 5465 | | 13 | 10 | - | 6 | 2 | - | 2 | 1 | - |

Table 2: Per-component Cost Breakdown. Units are in seconds, unless otherwise noted.

| Log Op. | bool | Int (4) | Address (8) | Str. | Array |
|---|---|---|---|---|---|
| Lat. (ns) | 34 | 35 | 39 | 158 | 278 |

Table 4: Latencies of Agent Operations. Address is a long integer for the address of a Java object retrived from `System.identityHashCode()`. Str. is a Java string of length 10. Array is a byte array of length 32.

### 6.2 Microbenchmarks

**Per-component Costs.** We measure the time spent on each phase of Lumos. Table 2 summarizes the results. The analysis time includes the time for running static analysis and that for SDG construction. The result illustrates the scalability of Lumos' analysis as the time is moderate. Next, the time spent on the controller's graph traversal in each debugging round (Rd.) is low except for L4022-1. We believe a highly optimized traversal algorithm can further improve this latency. Combined with the low instrumentation time (Inst.) for the `NameNode` and the `DataNodes`, we conclude that Lumos could swiftly respond to debugger's queries.

In addition, we measure the average latencies of the agent operations. This includes `startTrace`,`endTrace`, `setCaller`, `getCaller` and log operations that each carries a payload of one of the 5 representative content types used in our case studies. As shown in Table 4, Lumos agents' operations are lightweight.

## 7 Limitations

In this section, we discuss the limitation of Lumos. First, Lumos only records application-space values and provenances and does not cover libraries and lower layers of the system stack (e.g., operating system). We leave the work of cross-layer provenance recording as future work.

Second, Lumos supports limited forms of shared-state dataflow tracing. It only tracks accesses to thread-shared data structures. We leave supporting others (e.g., shared state in file systems) as future work. Furthermore, Lumos approximates orderings of concurrent events, which may be ineffective for concurrency bugs that require finer time resolution to resolve. We consider such concurrency bugs as an orthogonal direction [23], as Lumos' goal is to help developers link bugs' manifestation to their root causes.

Third, Lumos currently requires extra manual work between each debugging round (e.g., inferring values from captured non-determinism). We leave building an automatic provenance reconstructor as future work.

Finally, Lumos relies on a small set of developer annotations to track library-induced dependencies. When a system uses a large number of libraries exposing complex external semantics, it may be hard for Lumos to strike a good trade-off between the amount of manual labor and analysis precision. However, we believe recent trends in automatic library semantic summarization may alleviate this problem [12, 20, 38].

## 8 Related Work

**Record-and-Replay.** Record-and-replay-style tools capture all sources of non-determinism at runtime to ensure faithful replay of any execution trace a debugger to step through [27, 28, 31, 32]. Traditional RR tools typically incur high runtime overheads. Lumos avoids high recording overheads by applying static analysis to selectively record bug-provenance-related infromation only and using efficient shared-state tracing techniques.

**Automatic Logging & Tracing.** [42, 46] performs static analysis to add logs for reducing control- and data- flow ambiguities in buggy executions. However, they miss important dependencies such as inter-thread and inter-prodecedural heap data dependencies. As a result, they cannot recover deep chains of provenances to locate latent root causes. Distributed tracing tools [1, 8] automatically instrument tracepoints at predefined locations. However, the instrumentation is coarse and cannot recover evidence in the forms of variables' values. Lumos tracks all application-space dependencies and dynamically instruments logs to recover key values and provenances.

**Debugging In-producetion Bugs.** Several tools have been developed to troubleshoot in-production bugs. Gist [22] employs program slicing and hardware tracing to record bug-relevant data after a runtime incident occurs. However, its usage of hardware tracepoints cannot resolve heap dataflows involving addresses not known statically. Therefore, it cannot recover long chains of provenances involving heap dataflows. Failure reproduction tools [45, 47] synthesize bug-triggering execution via symbolic execution. Yet the tools may fail to reproduce complex bugs due to scalability issues and cannot guarantee identical root causes. Lumos uses scalable static analysis to guide fidel recording of actual runtime data.

## 9 Conclusion

Lumos enables lightweight and fast provenance-guided online debugging for distributed systems through automatic dependency-guided instrumentation of logs and tracepoints.

Lumos achieves this through scalable static analysis that guides efficient tracing runtimes to capture coherent provenances of bugs. Our experiences showed that Lumos could support debugging complex bugs with latent root causes spanning multiple components and traces with low diagnosis latency and minimal overheads. We believe that Lumos can open up a new revenue for efficient runtime debugging of complicated incidents in real-world in-production distributed systems.